\documentclass[preprint2]{aastex}

\newcommand{\lae}{\mathrel{<\kern-1.0em\lower0.9ex\hbox{$\sim$}}}
\newcommand{\gae}{\mathrel{>\kern-1.0em\lower0.9ex\hbox{$\sim$}}}
 
\newcommand{\keV}{\,{\rm keV} }
\shorttitle{The High Energy Spectrum of NGC 4151}
\shortauthors{Beckmann et al.}

\begin{document}

\title{The High Energy Spectrum of NGC 4151}


\author{V. Beckmann\altaffilmark{1}, C. R. Shrader\altaffilmark{2}, N. Gehrels}
\affil{NASA Goddard Space Flight Center, Exploration of the Universe Division, Greenbelt, MD 20771, USA}
\email{beckmann@milkyway.gsfc.nasa.gov}
\author{S. Soldi\altaffilmark{3}}
\affil{INTEGRAL Science Data Centre, Chemin d' \'Ecogia 16, 1290
 Versoix, Switzerland}
\author{P. Lubi\'nski\altaffilmark{4}, A. A. Zdziarski}
\affil{Nicolaus Copernicus Astronomical Center, Bartycka 18, 00-716 Warsaw, Poland}
\author{P.-O. Petrucci}
\affil{Laboratoire d'Astrophysique de Grenoble, BP 53X, 38041 Grenoble Cedex, France}
\and
\author{J. Malzac}   
\affil{Centre d'\'Etude Spatiale des Rayonnements, 31028 Toulouse, France}


\altaffiltext{1}{also with the Joint Center for Astrophysics, Department of Physics, University of Maryland, Baltimore County, MD 21250, USA}
\altaffiltext{2}{also with Universities Space Research Association, 10211 Wincopin Circle, Columbia, MD 21044, USA}
\altaffiltext{3}{also with the Observatoire de Gen\`eve, 51 Ch. des Maillettes, 1290 Sauverny, Switzerland}
\altaffiltext{4}{also with the INTEGRAL Science Data Centre, Chemin d' \'Ecogia 16, 1290 Versoix, Switzerland}

\begin{abstract}
We present the first {\it INTEGRAL} observations of the type 1.5 Seyfert galaxy NGC 4151. Combining several {\it INTEGRAL} observations performed during 2003, totaling $\sim 400 \, \rm ks$ of exposure time, allows us to study the spectrum in the 2 -- 300 keV range. 
The measurements presented here reveal an overall spectrum from X-rays up to soft gamma-rays that can be described by an absorbed ($N_{\rm H} = 6.9 \times 10^{22} \rm \, cm^{-2}$) model based on a Compton continuum from a hot electron population ($kT_e = 94 \keV$) from an optically thick ($\tau = 1.3$) corona, reflected on cold material ($R=0.7$), consistent with earlier claims. 
The time resolved analysis shows little variation of the spectral parameters over the duration of the {\it INTEGRAL} observations. The comparison with {\it CGRO}/OSSE data shows that the same spectral model can be applied over a time span of 15 years, with flux variations of the order of a factor of 2 and changes in the underlying continuum reflected by the temperature of the electron population ($kT_e = 50 - 100 \keV$). 
When modeled with an exponentially cut-off power law plus Compton reflection this results in photon indices ranging from $\Gamma = 1.5$ to $\Gamma = 1.9$ and a cut-off energy in the range $100 - 500 \keV$.
\end{abstract}




\section{Introduction}
The Seyfert galaxy NGC 4151 is one of the most extensively studied AGN (Active Galactic Nucleus). This is due to the fact that it is one of the brightest ($V \simeq 11 \rm \, mag$, bolometric luminosity $\sim 10^{44} \rm \, erg \, s^{-1}$) and nearest ($16.5 \rm \, Mpc$ for $H_0 = 75 \rm \, km \, s^{-1}$) among its class. Observations throughout the electromagnetic spectrum revealed a complex and somewhat special case of a Seyfert 1.5 galaxy (see e.g. Dermer \& Gehrels (1995) and Ulrich (2000) for a general review). In the optical the host galaxy appears to be a large barred spiral with nearly face-on orientation, showing two thin spiral arms \cite{arp77}.
At soft X-rays NGC 4151 {\it Chandra} data show a complex spectrum, including line emission from a highly ionized and variable absorber, in the form of H-like and He-like Mg, Si, and S lines, as well as lower ionization gas via the presence of inner-shell absorption lines from lower-ionization species of these elements \cite{NGC4151Chandra2}. Due to the heavy absorption, it is not possible to tightly constrain the intrinsic X-ray photon index below 10 keV. The harder continuum at $2 - 10 \rm \, keV$ as studied by {\it ASCA} shows a flatter spectrum with $\Gamma \sim 1.5$ \cite{NGC4151ASCA}. Observations by {\it XMM-Newton} in the $2.5 - 12 \rm \, keV$ energy range suggest that the spectrum can be modelled by an absorbed power-law continuum (fixed $\Gamma = 1.65$), a high energy break fixed at $100 \, \rm keV$, a neutral Compton reflection component with a reflection fraction of 2.0 compared to the normalization of the power law, and a narrow iron K$\alpha$ line. The fact that $R > 1$ is explained by the relative variability timescales in the direct and reflected component. The absorption of the power-law continuum was represented by a product of two absorption components \cite{schurch2}. The Galactic hydrogen column density in the line of sight is $N_{\rm H} \simeq 2.1 \times 10^{20} \, \rm cm^{-2}$.
The spectrum detected by {\it BeppoSAX} seemed to be best described by a power law with an exponential cut-off at $> 50 \rm \, keV$ and an iron fluorescence K$\alpha$ line 
\citep{piro99,piro00}. In addition to the fluorescence line, which reveals the existence of cooler material surrounding the AGN, a reflection hump in the $10-30 \rm \, keV$ has been reported in Ginga/OSSE data \cite{reflectionGinga}, in combined {\it ASCA} and OSSE data \cite{zdziarski}, and in {\it BeppoSAX} data
(Petrucci et al. 2001; Schurch \& Warwick 2002).

In this paper we present analysis of recent observations of NGC 4151 performed by the {\it INTEGRAL} ({\it International Gamma-Ray Astrophysics Laboratory}) satellite, and compare the results with previous studies.

\section{Simultaneous optical, X-ray and gamma-ray observations}

Observations in the X-ray to soft gamma-ray domain have been performed in late May 2003 by the {\it INTEGRAL} satellite \cite{INTEGRAL}. This mission offers the unique possibility to perform simultaneous observations over the $2 - 8000 \rm \, keV$ energy region. This is achieved by {\it INTEGRAL}'s Joint European X-ray Monitor (2--30 keV) (JEM-X; Lund et al. 2003), the {\it INTEGRAL} Soft Gamma-Ray Imager (20--1000 keV) (ISGRI; Lebrun et al. 2003), and the Spectrometer aboard {\it INTEGRAL} (SPI; Vedrenne et al. 2003), which operates in the 20 -- 8000 keV region.These three instruments use the coded aperture method for imaging. For a review on the particularities of this technique see Caroli et al. (1987).
In addition to these data an optical monitor (OMC, Mas-Hesse et al. 2003) provides $V$ band photometric measurements. 

In order to compare the {\it INTEGRAL} results with another simultaneous mission, we use data from the {\it Rossi X-ray Timing Explorer} ({\it RXTE}). 

\subsection{{\it INTEGRAL} data}

For the analysis of the high energy spectrum of NGC 4151 we used the data which were taken within 10 degrees radius around the source. Note that because of the nature of coded mask imaging, the whole sky image taken by the instrument has to be taken into account in the analysis, as all sources in the field of view contribute to the background \cite{codedmask}. A journal of our observations is shown in Table~\ref{journal}. The exposure are the ISGRI effective on-source times. This value is approximately the same for the spectrograph SPI, but the JEM-X and OMC monitors 
cover a much smaller sky area. Thus in the case of dithering observation, the source is not always in the field of view of the monitors. 
The analysis was performed using version 4.2 of the Offline Science Analysis (OSA) software distributed by the {\it INTEGRAL} Science Data Centre (ISDC; Courvoisier et al. 2003b).

The analysis of the {\it INTEGRAL}/IBIS data is based on a cross-correlation procedure between the recorded image on the detector plane and a decoding array derived from the mask pattern \cite{IBISOSA}.
Imaging analysis of the ISGRI data finds NGC 4151 at $\rm R.A. = 12^{\rm h} 10^{\rm m} 32^{\rm s}$, $\rm decl. = 39^{\circ} 24' 42''$ (J2000.0), $24''$ off the nominal optical position, with a significance of $224 \sigma$. The error radius of the position determination is $18''$. Note that this positioning accuracy is far below the instrumental angular resolution of $12'$ (see e.g. Gros et al. 2003).
The error radius rules out the two other X-ray sources within $10'$, the LINER NGC 4156 and the BL Lac object MS 1207.9+3945 (both at $\sim 5'$ distance to NGC 4151 and both not detected by {\it INTEGRAL}'s instruments), to be the counter part of the hard X-ray emission.

No cross-calibration correction had to be applied between JEM-X and ISGRI, but the SPI data appear to be at a higher level by a factor of 1.1. The SPI analysis was done using the specific analysis software \cite{SPIOSA} including version 9.2 of the reconstruction software SPIROS \cite{SPIROS} which is based on the ``Iterative Removal of Sources'' technique (IROS; Hammersley et al. 1992): a simple image of the field of view is made using a mapping technique which is optimised for finding a source assuming that the data can be explained by only that source plus background. The mapping gives the approximate location and intensity of the source, which are then improved by maximising a measure of the goodness of fit. The residuals of the fit are used as the input for a further image reconstruction and source search \cite{SPIROS}. The SPI data further include some energy bins where SPIROS could not find a solution, and others which indicate problems in the background subtraction. This is not surprising as most of the data have been taken in staring mode, which is known to cause problems with background determination in the SPI analysis \cite{Dubath}.

The ISGRI spectra have been extracted using the standard method as described in Goldwurm et al. (2003): fluxes and count spectra are extracted at the source positions for each pointing 
in predefined energy bins. Spectra of the same source collected in different pointings or ``science windows'' (SCWs) are then summed to obtain an average spectrum during the observation. Regarding the absolute flux calibration, it must be noted that the instrumental response of ISGRI is still not accurately determined. This problem is most severe for off-axis observations. In addition the spectra of the Crab taken with JEM-X show systematic features in the $5 - 7 \rm \, keV$ energy range, which might affect the measurement of the iron K$\alpha$ line. 

With respect to the calibration uncertainty, the IBIS instrument team stated that the systematic error is of the order of 1.5\% (2005, private communication). Nevertheless, this value corresponds to on-axis observations within a short period of time with no disturbing influence, such as enhanced background activity. A combined fit of Crab spectra taken in revolution 43, 44, 120, 170, and 239, i.e. over a 1.5 year span of the mission, shows a larger uncertainty in the flux. The scale of diversity, assuming that Crab is a source with constant flux, gives some hint what is the scale of
discrepancies in count rates observed in various conditions, i.e., with different
dithering patterns or instrumental settings. Clearly the 
absolute calibration cannot be more precise than the observed variations. Hence, it
seems that for the 22-120 keV band we may conclude that the joint $1 \sigma$ uncertainty of the spectral extraction method and the calibration files is about $5\%$. It may happen that for other sources and in some special circumstances
we may encounter a larger discrepancy, but for NGC 4151 observations after
revolution 65 and done mainly in staring mode a systematic error of 5\% should be a valid approximation.

The spectral shape calibration was also tested on {\it INTEGRAL} Crab observations. With the so-called canonical model for the Crab showing a single power law with photon index $\Gamma = 2.1$ \cite{Crab}, the values retrieved from {\it INTEGRAL} are $\Gamma = 2.1$ (JEM-X; Kirsch et al. 2004), and $\Gamma = 2.2$ for SPI and ISGRI. It should be kept in mind though that the Crab is significantly brighter than NGC 4151, and systematic effects might depend on source brightness.

NGC 4151 was too faint above $\sim 150 \rm \, keV$ to be detected by IBIS/PICsIT. Summing all available {\it INTEGRAL} data together shows that NGC 4151 is detected up to 300 keV, in a total exposure of 408 ks.

\subsection{{\it RXTE} data}
\begin{figure}
\plotone{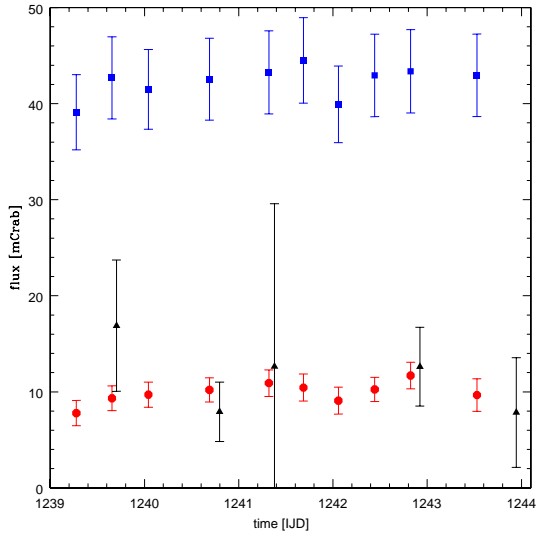}
\caption[]{{\it INTEGRAL} and {\it RXTE}/ASM light curves in Crab units. The squares and octagons represent the $20 - 100 \rm \, keV$ ISGRI and $1.5-12 \rm \, keV$ JEM-X2 lightcurve, respectively. Fluxes were extracted from model fits to the data. The triangles are {\it RXTE}/ASM $1.5 - 12 \rm \, keV$ daily averages.}

\label{fig:lightcurve}
\end{figure}

The {\it RXTE} All Sky Monitor (ASM; Levine et al. 1996) scans about 80\% of the sky every orbit. This offers a unique way to monitor the emission of bright X-ray sources like NGC 4151 in the $1.5 - 12 \rm \, keV$ energy range. We extracted fluxes from the {\it RXTE}/ASM data base. The fluxes have been averaged over one day (see the triangles in Fig.~\ref{fig:lightcurve}).
The data show a $10 \pm 3 \, \rm mCrab$ flux over the duration of the {\it INTEGRAL} observations, consistent with the JEM-X2 measurement.  

\subsection{{\it CGRO} data}

The source was also monitored in the $\sim 20 - 100 \rm \, keV$ band by the {\it CGRO}/BATSE experiment from 1991-2000 \citep{parsons,harmon}. Those measurements are made using an Earth-occultation technique, and were thus sampled one time per 90 min {\it CGRO} orbit. In practice, NGC 4151 is a faint source for that instrument, but daily summations lead to $\sim 3 \sigma$ determinations. We obtained the BATSE data spanning the $\sim 3.5 \rm \, yr$ overlap of the {\it RXTE} and {\it CGRO} missions, for the purpose of monitoring the long-term hard-to-soft spectral behavior. 

\section{X-ray to Gamma-ray spectrum}

\subsection{Time Averaged Spectral Analysis}

\begin{figure}
\plotone{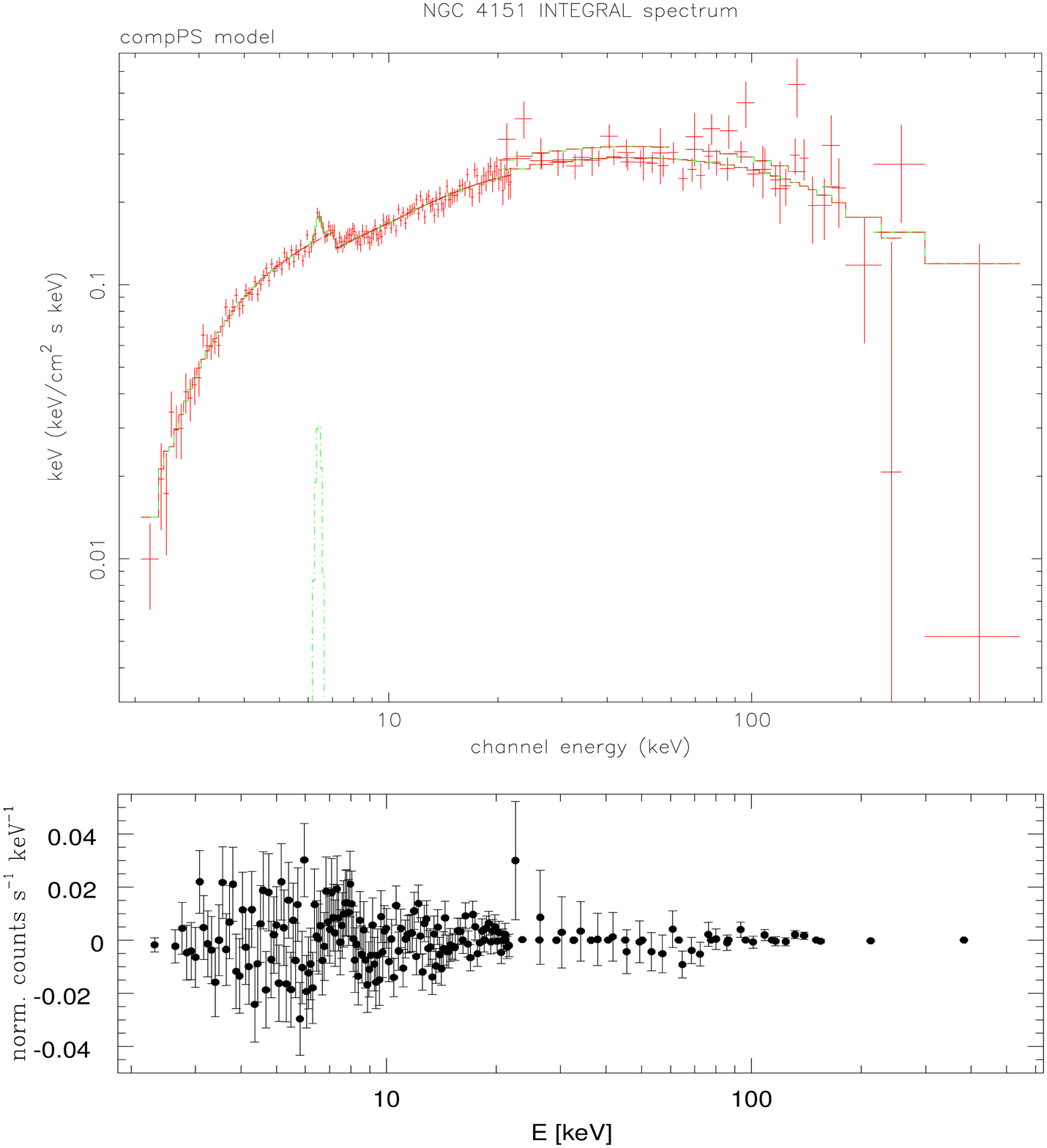}
\caption[]{Summed spectrum of the {\it INTEGRAL} JEM-X2, ISGRI, and SPI data ({\it upper panel}). The data are well represented by the Compton reflection model {\tt compPS} with electron temperature of  $kT_e = 94 \keV$ and relative reflection of $R = 0.7$ . In addition a narrow gaussian iron K$\alpha$ line with equivalent width $EW = 119 \rm \, eV$ appears in the spectrum. The lower panel shows the residuals of the fit. }
\label{fig:combinedspec}
\end{figure}

The overall spectrum extracted from the {\it INTEGRAL} JEM-X, ISGRI, and SPI data is shown in Fig.~\ref{fig:combinedspec} together with the residuals of the model fit. A systematic error of 5\% has been added to the ISGRI and SPI data. The model fitting results however do not depend strongly on the SPI data, because of the larger error bars compared to the ISGRI data.

The spectral fitting was done using version 11.3.2 of XSPEC \cite{XSPEC}. The data cover the energy range $2 - 300 \rm \, keV$. We applied a fit of an absorbed power law with an exponential high-energy cut-off, and added a Gaussian line in order to account for the iron K$\alpha$ fluorescence line. The best fit values are $\Gamma = 1.53 {+0.04 \atop -0.04}$, $E_C = 124 {+21 \atop -16} \rm \, keV$, and $N_{\rm H} = 5.1 {+0.7 \atop -0.7} \times 10^{22} \rm \, cm^{-2}$. The fluorescence line we obtain has an equivalent width of $EW = 119 \rm \, eV$ and a flux of $f_{\rm K\alpha} = 4.5 {+1.3 \atop -1.6} \times 10^{-4} \rm \, ph \, cm^{-2} \, s^{-1}$ (all errors are 90\% confidence values). The statistics do not allow us to assess possible effects of relativistic broadening of the iron line (see e.g. Yaqoob et al. 1995). The source had an average flux of $f_{20-200 \rm \, keV} = 0.012 \rm \, ph \, cm^{-2} \, s^{-1} = 9.6 \times 10^{-10} \rm \, erg \, cm^{-2} \, s^{-1}$ during the observation. This model gives $\chi^2_\nu = 1.22$ for 196 degrees of freedom. 
The model of an absorbed power law without a cut-off leads to a significantly poorer fit result ($\chi^2_\nu = 2.1$). 

A physical model describing 
Comptonization of soft photons by a hot plasma, the so called {\tt compTT} model, has been developed by Titarchuk (1994). The model includes the plasma temperature $T_e$ of the hot corona, the optical depth $\tau$ of this plasma, and the temperature $T_0$ of the soft photon spectrum. Because the spectrum starts at 2 keV, $T_0$ is not well constrained by the data and has been fixed to 10 eV. The model fit resulted in $N_{\rm H} = 6.4 {+0.5 \atop -0.5} \times 10^{22} \rm \, cm^{-2}$, $k T_e = 39 {+11 \atop -7} \, \rm keV$, and $\tau = 1.37 {+0.26 \atop -0.33}$ for plane geometry. The {\tt compTT} model gives $\chi^2_\nu = 1.29$ for 195 degrees of freedom.

Because those models did not give conclusive fit results, we added a reflection component from cold material to the cut-off power-law (the so-called PEXRAV model; Magdziarz \& Zdziarski 1995). 
From radio measurements the inclination angle $i$ of the accretion disk is estimated to be $i \sim 65^\circ$ \cite{inclination}, where $i = 0$ would be the case of a face-on disk. The free parameters in the PEXRAV model are best fitted with a photon index of $\Gamma = 1.85 {+0.09 \atop -0.09}$, a hydrogen column density of $N_{\rm H} = 6.9 {+0.8 \atop -0.4} \times 10^{22} \rm \, cm^{-2}$, a cut-off energy of $E_{C} = 447 {+885 \atop -190} \rm \, keV$ and a relative reflection of $R = 1.0 {+0.4 \atop -0.3}$. 
The large uncertainty and high value of the cut-off energy, which is well outside the energy range covered by our detection, indicates that this parameter is not well constrained. The PEXRAV model leads to $\chi^2_\nu = 0.99$ for 195 degrees of freedom.


Although this model is a good representation of the data, the choice of a power law for the underlying high-energy continuum is purely phenomenological. A more physically approach involves the {\tt compPS} model \cite{compPS}. This is a self-consistent model which computes a Comptonization spectra using exact numerical solutions of the radiative transfer equation. The resulting spectrum is reflected from the cool medium according to the computational method of Magdziarz \& Zdziarski (1995). The model fit resulted in an electron temperature of $kT_e = 94 {+4 \atop -10} \keV$, an optical depth of the corona $\tau = 1.31 {+0.15 \atop -0.05}$, and a relative reflection of $R = 0.72 {+0.14 \atop -0.14}$, with $\chi^2_\nu = 0.98$ for 196 degrees of freedom. 

We also applied models used in previous studies. For example, using the model described in Schurch et al.~(2003) with the continuum fixed to an absorbed power law with $\Gamma = 1.65$, the fit result worsens to $\chi^2_\nu = 1.06$, and when also fixing the high energy break to $100 \rm \, keV$ the fit results in $\chi^2_\nu = 1.39$. It should be pointed out that though Schurch et al. used {\it ASCA}, {\it BeppoSAX}, and {\it XMM-Newton} data to develop their spectral 'template' model, and these data do not constrain a high-energy cut-off above 100 keV particularly well.

Using different strength of the reflection parameter, as described in Petrucci et al. (2001) also led to poorer fit results. For $R = [0.0, 0.01, 0.2, 0.45]$ the fit resulted in $\chi^2_\nu = [1.22, 1.21, 1.11, 1.03]$. A reflection component with $R > 0.5$ is therefore required to achieve an acceptable fit result to the {\it INTEGRAL} data.

Although previous studies showed that the spatial and dynamical structure of the absorbing media in NGC 4151 is more complex 
(Kraemer et al. 2005; Ogle et al. 2000; Schurch \& Warwick 2002; Schurch et al. 2003)
than the single cold absorber applied in our model, and that the soft X-rays are dominated by radiative recombination continua and X-ray emission lines \cite{NGC4151Chandra2, schurch3}, a more complicated model incorporating such would not improve the fit results to the {\it INTEGRAL} data. This is due to the fact that the JEM-X data in the relevant range up to $\sim 3 \rm \, keV$ do not have sufficiently high signal-to-noise to derive constraining information from this part of the spectrum.

\subsection{Spectral Evolution}

\begin{figure}
\plotone{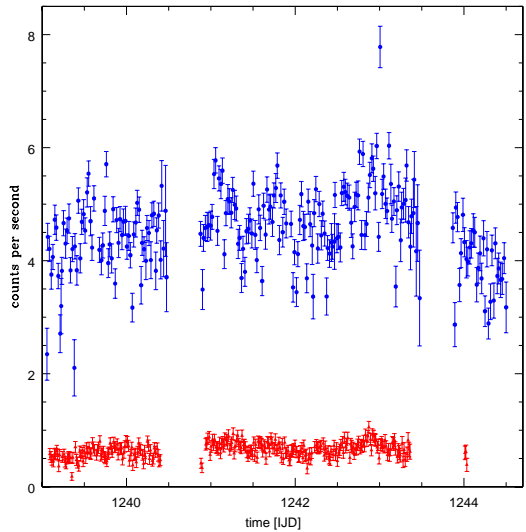}
\caption[]{{\it INTEGRAL} ISGRI ({\it upper curve}, 2000 s binning) and JEM-X2 (1000 s binning) lightcurve in the $20 - 100 \rm \, keV$ and $5-20 \rm \, keV$ energy band, respectively.}
\label{fig:ISGRIJEMXlightcurve}
\end{figure}

In order to study the evolution of the spectrum in time, we applied ten time bins over the total observation period, combining about $33 \rm \, ks$ of effective on-time in each bin. We used only those data for which we have simultaneous JEM-X and ISGRI coverage. We then applied the cutoff power-law model to the individual time bins for ISGRI and an absorbed power law model for JEM-X and extracted fluxes in the $20 - 100 \rm \, keV$ and $1.5 - 12 \rm \, keV$ range, respectively. The results are shown in Fig.~\ref{fig:lightcurve} together with the {\it RXTE}/ASM lightcurve in the same energy band as JEM-X. No significant variation of the flux can be seen here, and also the fit parameters are consistent over the observed period within the error bars. 
In order to look for variability on shorter time scales we then binned the {\it INTEGRAL} JEM-X2 (5 -- 20 keV) and ISGRI (20 -- 100 keV) data into 2000 s time intervals (Fig.~\ref{fig:ISGRIJEMXlightcurve}). Both lightcurves show an apparent flare at $t \simeq 1243.0 \rm \, IJD$\footnote{IJD is related to the  modified Julian Date by $IJD = MJD - 51544.0$}. A more careful analysis of the data enclosed within that time bin, $1242.5 {\rm \, IJD} < t < 1243.5 {\rm \, IJD}$, does show a slightly increased flux ($f_{20-200 \rm \, keV} = 0.0132  \rm \, ph \, cm^{-2} \, s^{-1} = 1.03 \times 10^{-9} \rm \, erg \, cm^{-2} \, s^{-1}$) but the spectral parameters are consistent with the time-averaged values.

We performed a cross-corrleation analysis of the JEM-X and ISGRI light 
curves (Fig.~\ref{fig:ISGRIJEMXlightcurve}), applying the "Discrete Cross-Correlation" algorithm 
of Edelson \& Krolik (1988). A lag-spectrum spanning 
$\pm 0.5$ days, with a granularity of 0.05 days (or roughly 2.5 times our 
sampling time) was constructed. A plot of the resulting correlation statistic versus lag 
was symmetric about 0, with a peak value of about 40 \%. This suggests that the hard and soft X-ray variations are well correlated, but that our sampling timescales are too coarse to study any reprocessing timescale on or light travel-time effects.

\begin{figure}
\plotone{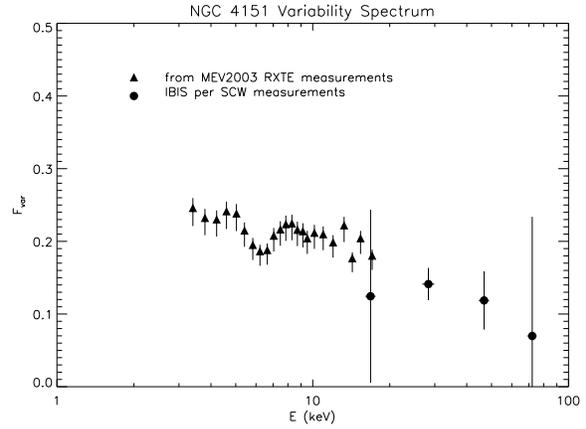}
\caption[]{Fractional rms variability amplitude based on the {\it INTEGRAL} ISGRI count rates ({\it circles}) compared to the results of Markowitz, Edelson, \& Vaughan (2003; {\it triangles}).}
\label{fig:varspec}
\end{figure}

We also evaluated the hard-X-ray spectral variability following the procedure 
of Vaughan et al. (2003) and Markowitz, Edelson, \& Vaughan (2003). We computed the fractional rms variability amplitude based on the per-SCW count rates and uncertainties on each of 4 energy bands: 13-20, 20-36, 
36-58, and 58-87 keV. Our results are limited by the data; there were a 
total of 244 SCWs spanning about 7 days. The signal-to-noise ratios 
ranged from $\sim 17$ in the best case (20-36 keV) to $\sim 3-5$ in the poorest case (13-20 keV). For purpose of comparison, we have reproduced the lower 
energy results of Markowitz, Edelson, \& Vaughan (2003), and presented with our results in Fig.~\ref{fig:varspec}. 

\begin{figure}
\plotone{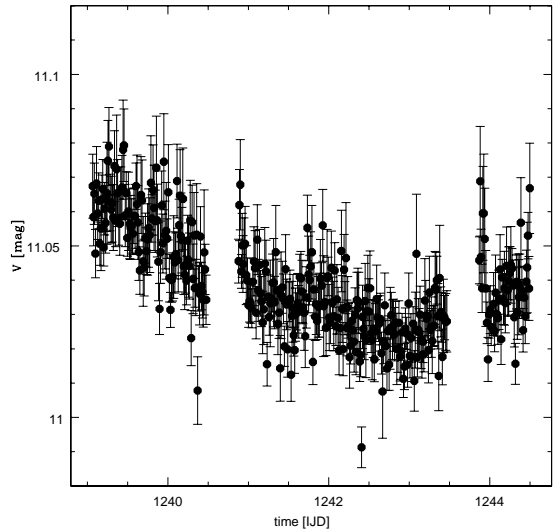}
\caption[]{{\it INTEGRAL}/OMC light curve. Error bars are $1 \sigma$ values.}
\label{fig:OMClightcurve}
\end{figure}

For the same time intervals as used in the high-energy analysis, optical photometry values from the OMC have been extracted. Due to the high signal-to-noise these data have been binned into $1000 \rm \, s$ time intervals (Fig.~\ref{fig:OMClightcurve}). 
We also considered possible correlation between variations in our 
optical measurements and the ISGRI hard X-ray rates. We derived the 
approximate mono-chromtatic flux ($f_{5500 \rm \, \AA}$) from the OMC V-band 
magnitudes, and normalized these values to $10^{-12} \rm \, ergs \, cm^{-2} \, s^{-1} \, \AA^{-1}$. We then 
applied the same cross-correlation algorithm to the X-ray count rates and the 
optical flux. The same range and granularity as we used for the JEM-X - 
ISGRI case was applied. The results obtained indicate a lack of 
significant correlation (peak amplitude $\sim 25 \%$), and a flat, symmetric 
distribution about zero-lag.

\subsection{Search for $\gamma$-ray line emission}

The SPI spectrometer offers the opportunity to look for narrow line features. In case the high energy spectrum would be dominated by non-thermal processes, one could expect a significant amount of $\rm e^\pm$ pair annihilation processes around $511 \rm \, keV$. At this energy, SPI has a resolution of $FWHM = 1.95 \rm \, keV$ \cite{attie}. {\it CGRO}/OSSE was able to determine the $3 \sigma$ upper limit of a broad ($450 - 600 \rm \, keV$) emission line feature with $f < 6 \times 10^{-5} \rm \, ph \, cm^{-2} \, s^{-1}$. We therefore looked for a comparably narrow feature by extracting an image in the $499 - 519 \rm \, keV$ energy band, centred on the 511 keV rest-frame energy at NGC 4151's redshift. No emission at the restframe energy of the annihilation line is detectable. The $3 \sigma$ upper limit extracted from the SPI data by SPIROS is $f_{499 - 519 keV} = 10^{-4} \rm \, ph \, cm^{-2} \, s^{-1}$. The $3 \sigma$ line sensitivity derived from ground calibration gives a similar value of $6.5 \times 10^{-5} \rm \, ph \, cm^{-2} \, s^{-1}$ \cite{roques}. The measurement presented here is therefore not improving the {\it CGRO}/OSSE upper limit, because of the much shorter {\it INTEGRAL}/SPI exposure time.

\section{Discussion}

The overall {\it INTEGRAL} spectrum can be modelled by Comptonization of photons in a hot plasma, and suggest that such an environment characterizes the central regions of NGC 4151.

\begin{figure}
\plotone{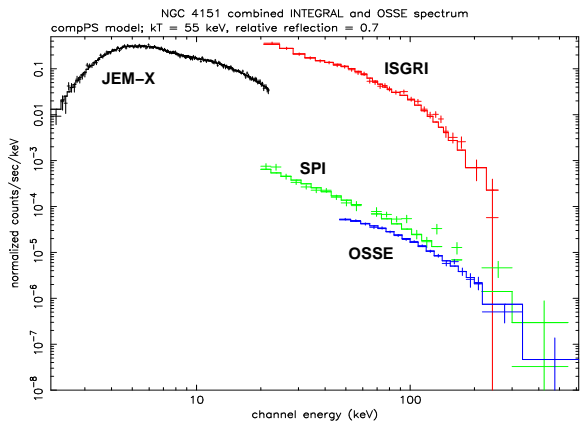}
\caption[]{{\it INTEGRAL} and {\it CGRO}/OSSE simultaneous data fit. The OSSE data represent the average of the NGC 4151 spectra. All data are well represented by a Compton reflection model ({\tt compPS}; Poutanen \& Svensson 1996), with the individual OSSE data scaling with $0.6 \dots 1.2$ versus the {\it INTEGRAL}/JEM-X2 data.}
\label{fig:INTOSSE}
\end{figure}

The best comparable coverage in the energy range studied by {\it INTEGRAL} ISGRI and SPI has been provided by {\it CGRO}/OSSE \cite{OSSE4151}. The measurements from {\it INTEGRAL} and OSSE are compared in Tab.~\ref{modelfit}. The OSSE results have been derived by fitting simultaneously the individual OSSE spectra, allowing for different normalisation. Our results are similar to what has been reported in Johnson et al. (1997).
We compared the OSSE data directly with the {\it INTEGRAL} observation through simultaneous fitting. Figure~\ref{fig:INTOSSE} shows a simultaneous fit of both data sets, the average {\it INTEGRAL} and average {\it CGRO}/OSSE spectrum. The data can be represented by a (non-physical) cutoff power law with individual constant factors in order to account for flux variability ($\chi^2_\nu = 1.10$). Note that the energy range above $\sim 100 \rm \, keV$ is dominated by the OSSE data, as their error bars are significantly smaller than the {\it INTEGRAL} ones, because the {\it INTEGRAL} exposure time is much smaller. The fit shows that the $20 - 200 \rm \, keV$ flux varied by a factor of 2 over the individual {\it INTEGRAL} and {\it CGRO} observations. It is remarkable that the same simple model with $\Gamma = 1.5$ and $E_C = 95 \rm \, keV$ can represent the data from the different {\it CGRO}/OSSE and {\it INTEGRAL} observations.

When applying the more meaningful PEXRAV or compPS model to the {\it CGRO} data it turns out that a reflection component is not constrained by the OSSE data. This is not surprising, as there are no OSSE data below 50 keV where a reflection component would have a significant influence. The PEXRAV fit to the combined data shows that a reflection component improves the fit according to an F-test. The reduced $\chi^2$ is 1.07 and the relative reflection is only $R = 0.4 {+0.2 \atop -0.1}$ in this case.
The compPS model gives a better fit result ($\chi^2_\nu = 1.01$ for 476 d.o.f.) with an electron temperature of $kT_e = 55 {+2 \atop -1} \keV$ and a relative reflection  of $R = 0.7 {+0.1 \atop -0.1}$. 

These results are mainly consistent with previous {\it BeppoSAX} observations. Three observations by {\it BeppoSAX}, two in 1996 and one in 1999, suggest that the strength of the reflection component is variable, with $R_{1996H}=0.01, R_{1999}=0.2$, and $R_{1996L}=0.45$ \cite{POP}. $R$ is normalized so that $R=1$ in the case of an isotropic source above an infinite reflecting plane. $R$ is often considered as an estimate of $\Omega/2\pi$ where $\Omega$ is the solid angle subtented by the reflector as seen from the isotropic X-ray source. It has to be pointed out that a relative reflection of $R > 1$, as seen in the work of Schurch et al. (2003), represents the case when more 
primary X-ray radiation is emitted toward the reflector than toward the observer.
This can be explained by variable emission of the central engine and a time delay between the underlying continuum and the reflected component, caused by a large distance of the reflecting material to the primary source \cite{malzac}. Another model which can lead to large $R$ values is based on relativistic effects caused by a dynamic corona moving towards the reflecting disk \cite{dynamiccorona}. Other explanations include a special geometry with a high intrinsic covering fraction of the cold disk material \cite{julienalone} and general relativistic light bending effects caused by a maximally rotating Kerr black hole \cite{lightbending}.


Another study of the high energy spectrum of NGC 4151 combining {\it CGRO}/OSSE and ASCA data has been presented by Zdziarski et al. (2002). This work came to the conclusion that the data can be represented by a simple cut-off power law with a narrow iron K$\alpha$ line, adding thermal bremsstrahlung at low energies, which are not covered by our data. This fit resulted in a $\chi^2_\nu = 1.6$ and the authors added a broad Gaussian line in order to account for residuals in the $5 - 6.5 \rm \, keV$ energy band. Zdziarski et al. argue that a model with complex absorption and Compton reflection component is more likely to give a physical meaningful explanation for the observed spectrum. 
They also discovered that the absorber in NGC 4151 shows evidence for the presence of several absorbing components in the line of sight, which cannot be confirmed by the {\it INTEGRAL} observations because of the lack of data below 2 keV.

Matt et al.~(2003) argue that absorption in Seyfert 2 galaxies can be caused by several circumnuclear regions. They favour a model with a possible temporary switching-off of the nuclear radiation to explain observed variability.
Their model predicts an iron K$\alpha$ equivalent width of $\sim 100 \, \rm eV$ for a Compton thick case with $N_{\rm H} \simeq 10^{23} \, \rm cm^{-2}$, consistent with the measurements for NGC 4151. This could mean that the iron line is produced in material in the line of sight, at large distances from the central engine.

Schurch \& Warwick (2002) showed that the same 'template' spectrum matches observations of NGC 4151 done by ASCA, BeppoSAX and even XMM-Newton \cite{schurch2}. Their model spectrum includes a strong reflection component ($R \simeq 2$ from PEXRAV) as well as an absorption component with variable iron abundance. 
The {\it INTEGRAL} data are not suitable for detailed study of the absorption components. The observations show that the underlying continuum is significantly steeper, with $\Gamma = 1.85 {+0.09 \atop -0.09}$ instead of $\Gamma = 1.65 {+0.02 \atop -0.03}$ \cite{schurch1}, and with a higher cut-off energy, although the fit to the {\it INTEGRAL} data using $\Gamma = 1.65$ is still acceptable ($\chi^2_\nu = 1.06$ instead of 0.99).
This shows that it is possible to represent the spectrum of NGC~4151 over a long time scale with the same model with only a few parameters showing significant changes.  
The PEXRAV model appears to be a good representation of the high energy spectrum of NGC 4151, with the continuum's photon index varying in the range $\Gamma = 1.5 \dots 1.9$, and the cut-off energy is $E_{C} > 90 \keV$ with no clear upper limit detected at present.
 

The variability study we performed on the {\it INTEGRAL} data showed that the hard and soft X-ray variations are well correlated during the observation.
The fractional rms variability amplitude based on the per-SCW count rates and uncertainties in four energy bands in comparison to an earlier study by Markowitz et al. (2003) suggests that there is a continuation of the weak anti-correlation, or even a flattening, of the variability with energy (Fig.~\ref{fig:varspec}).While one must bear in mind that these data are from separate epochs, 
they are consistent with a scenario where the putative Compton-reflected 
component is slowly varying relative to the disk emission. This in turn 
is suggestive of a reflector which may be larger in scale and/or farther 
from the low-energy photon sources than the inner accretion disk; perhaps 
associated with an outer disk torus, or with outflowing plasma.

\begin{figure}
\plotone{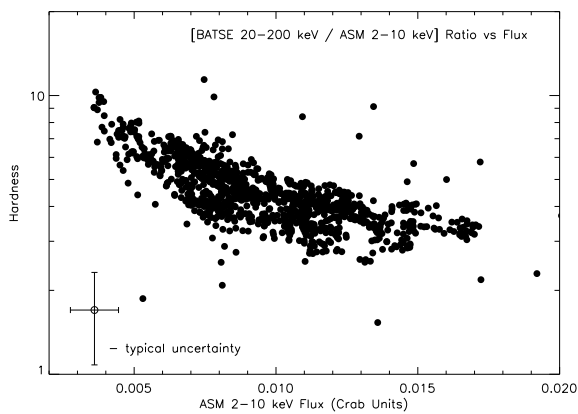}
\caption[]{Spectral hardness derived from the BATSE and RXTE data, versus the {\it RXTE}/ASM flux. A softer spectrum is correlated with a higher flux. The error bars on the lower left show the typical uncertainties of the measurements.}
\label{fig:BATSERXTE}
\end{figure}

Most Seyfert galaxies show a softening of the X-ray continuum as sources brighten (Markowitz, Edelson, \& Vaughan 2003; Nandra et al. 1997).
We did not see significant variations of the spectral parameters over the duration of our observations. In particular, when modelling the spectrum with a cut-off power-law, we did not detect any correlation of flux with spectral slope. This can be understood when studying the correlation of spectral hardness ratio with flux as derived from simultaneous {\it CGRO}/BATSE and {\it RXTE}/ASM data as shown in Fig.~\ref{fig:BATSERXTE}. Here we have plotted the hardness ratio $(20-200 \rm \, keV)/(2-10 \rm \, keV)$ versus the RXTE $(2-12 \rm \, keV)$ flux based on daily averages from 1996-2000. An overall correlation is discernible, but the scatter is large and thus might not be detectable over a time span of 400 ks (as in our case), and can even be missed when comparing the long-term observations by {\it CGRO}/OSSE. 



\section{Conclusion}

The results from the {\it INTEGRAL} JEM-X2, ISGRI, and SPI data show that the high-energy spectrum of NGC 4151 can be approximated by a rather simple model including Compton reflection.  
A Compton reflection model (PEXRAV) 
results in an underlying power-law with $\Gamma = 1.85 {+0.09 \atop -0.09}$, a hydrogen column density of $N_{\rm H} = 6.9 {+0.8 \atop -0.4} \times 10^{22} \rm \, cm^{-2}$, a cut-off energy of $E_{C} = 450 {+900 \atop -200} \rm \, keV$, and a relative reflection of $R = 1.0 {+0.4 \atop -0.3}$. 
However, the high-energy cut-off is still not well constrained by the data.
The more physical {\tt compPS} model reveals a hot electron population ($kT_e = 94 {+4 \atop -10} \keV$), an optically thick corona ($\tau = 1.3 {+0.2 \atop -0.1}$), and a significant reflection component ($R = 0.7 {+0.1 \atop -0.1}$). Even though the data do not constrain the reflection model strongly, they show the necessity of a reflection component.  
The study shows that not only the reflection component varies on long time scales, but that also the underlying continuum changes significantly over the years.

The comparison to previous {\it CGRO}/OSSE measurements shows that 
this model is a valid representation of the high energy spectrum over decadal timescales, with the normalisation varying by a factor of 2 and the temperature of the underlying continuum varying in the range $kT_e = 50 - 100 \keV$ for the various observations. A significant variation of the spectral parameters during the {\it INTEGRAL} observation cannot be detected. The effect of spectral softening with higher flux values can only be seen in observations covering longer time scales. This is shown in the comparison of {\it CGRO}/BATSE (20--200 keV) to {\it RXTE}/ASM (2--10 keV) fluxes.  

The {\it INTEGRAL} data cannot confirm the existence of a high iron overabundance in the absorber, simply because of the lack of data below $2 \rm \, keV$.

The thermal Comptonization is seen in the cut-off of the spectrum, ruling out synchrotron emission being dominant in this source. This is also supported by the non-detection of emission in the annihilation line in NGC 4151.

The study of NGC 4151 indeed shows that the {\it INTEGRAL} mission can give valuable insights on the characteristics of the high-energy cut-off in bright AGN, as has been already shown for the Seyfert 2 NGC 4388 (Beckmann et al. 2004a, 2004b). This shows that AGN can be well described by an optical thick disk with a hot corona. We will present a general overview on the properties of AGN observed by {\it INTEGRAL} in a forthcoming paper \cite{INTEGRALAGN}. 
\begin{acknowledgements}
We like to thank the referee for the comments which helped to improve the paper. 
This research has made use of the NASA/IPAC Extragalactic Database (NED) which is operated by the Jet Propulsion Laboratory and of data obtained from the High Energy Astrophysics Science Archive Research Center (HEASARC), provided by NASA's Goddard Space Flight Center. PL and AAZ were supported by KBN grants 1P03D01827, 1P03D01727,
4T12E04727 and PBZ-KBN-054/P03/2001. POP and AAZ were supported by the Polish-French program Astro-PF.
\end{acknowledgements}

%
%
%
\clearpage
%
%
%
\begin{table*}
\caption[]{{\it INTEGRAL} observations}
\begin{tabular}{lcrrc}
\tableline\tableline
obs. date & {\it INTEGRAL}   & exposure             & number of & observation \\
start     & revolution & time [ks] & pointings & mode        \\
\hline
 22/05/03 & 73         &       7.3      & 3         & $5 \times 5$\\
 23/05/03 & 74         &     170.8      & 10        & $5 \times 5$, staring\\
 25/05/03 & 75         &     174.5       & 5         & staring     \\
 28/05/03 & 76         &      55.1       & 4         & staring\\
\tableline
\end{tabular}
\label{journal}
\end{table*}
\begin{table*}
\caption[]{Spectral fits to average {\it INTEGRAL} (2 -- 300 keV) and OSSE (50 -- 600 keV) spectra}
\begin{tabular}{lccccc}   
\tableline\tableline
Model & Mission & Photon & $E_C$ or $kT$ & $\tau$ or $R$ & $\chi_{\nu}^2 (dof)$\\
    &        & Index  & $[\rm keV]$ & & \\  
\hline
Power law & {\it INTEGRAL}  & $1.79 \pm 0.02$ & $\dots$ & $\dots$ & 2.10 (197)\\
Power law & {\it CGRO}/OSSE & 2.55 $+0.03 \atop -0.03$           & $\dots$ & $\dots$ & 1.73 (278)\\
PL-exp    & {\it INTEGRAL}  & 1.53 $+0.04 \atop -0.04$ & 124 $+21 \atop -16$ & $\dots$ & 1.22 (196)\\
PL-exp    & {\it CGRO}/OSSE & 1.47 $+0.15 \atop -0.16$ &  90 $+16 \atop -12$ & $\dots$ & 0.98 (277)\\
PL-exp    &  {\it INTEGRAL}+{\it CGRO}& 1.49 $+0.04 \atop -0.04$ &  94 $+5 \atop -5$ & $\dots$ & 1.10 (476)\\
compTT$^a$ & {\it INTEGRAL}  & $\dots$ & 39 $+11 \atop -7$ & 1.37 $+0.26 \atop -0.33$ & 1.29 (195)\\
compTT$^a$ & {\it CGRO}/OSSE  & $\dots$ & 44 $+6 \atop -5$ & 0.89 $+0.18 \atop -0.17$ & 0.99 (276)\\
PEXRAV & {\it INTEGRAL} & 1.85 $+0.09 \atop -0.09$ & 447 $+885 \atop -190$ & 0.98 $+0.36 \atop -0.30$ & 0.99 (195)\\
PEXRAV & {\it CGRO}/OSSE & 1.52 $+0.25 \atop -0.28$ & 94  $+30 \atop -9$ & 0.00 $+2.78 \atop -0.00$ & 0.99 (276)\\
PEXRAV & {\it INTEGRAL}+{\it CGRO} & 1.61 $+0.06 \atop -0.06$ & 122 $+18  \atop -14$ & 0.40 $+0.18 \atop -0.08$ & 1.07 (475)\\
compPS$^{b}$ & {\it INTEGRAL} & $\dots$ & 94 $+4 \atop -10$ & $R= 0.72 {+0.14 \atop -0.14}$ & 0.98 (196)\\ 
compPS$^{b}$ & {\it CGRO}/OSSE & $\dots$ & 52 $+8 \atop -2$ & $R = 0.48 {+0.65 \atop -0.13}$ & 0.98 (277)\\ 
compPS$^{b}$ & {\it INTEGRAL}+{\it CGRO} & $\dots$ & 55 $+2 \atop -1$ & $R= 0.66 {+0.14 \atop -0.09}$ & 1.01 (476)\\ 
\tableline
\end{tabular}
\label{modelfit}

$^{a}$ Following Titarchuk 1994\\ 
$^{b}$ Following Poutanen \& Svensson 1995\\
\end{table*}

\end{document}